\documentclass[12pt]{article}
\usepackage{cite,graphicx}
\newcommand{\newc}{\newcommand}
\newc{\gsim}{\lower.7ex\hbox{$\;\stackrel{\textstyle>}{\sim}\;$}}
\newc{\lsim}{\lower.7ex\hbox{$\;\stackrel{\textstyle<}{\sim}\;$}}
\newc{\gev}{\,{\rm GeV}}
\newc{\mev}{\,{\rm MeV}}
\newc{\ev}{\,{\rm eV}}
\newc{\kev}{\,{\rm keV}}
\newc{\tev}{\,{\rm TeV}}
\def\ln{\mathop{\rm ln}}

\newc{\mz}{M_Z}
\newc{\mpl}{M_*}
\newc{\mw}{m_{\rm weak}}
\renewcommand{\epsilon}{\varepsilon}
\def\beq{\begin{equation}}
\def\eeq{\end{equation}}
\def\bea{\begin{eqnarray}}
\def\eea{\end{eqnarray}}
\newc{\ie}{{\it i.e.}}          \newc{\etal}{{\it et al.}}
\newc{\eg}{{\it e.g.}}          \newc{\etc}{{\it etc.}}
\newc{\cf}{{\it c.f.}}
\def\meff{m_{\rm eff}}
\def\bar#1{\overline{#1}}

\def\inv{^{\raise.15ex\hbox{${\scriptscriptstyle -}$}\kern-.05em 1}}
\def\lbar{{\lower.35ex\hbox{$\mathchar'26$}\mkern-10mu\lambda}} 

\def\to{\rightarrow}

\let\<=\langle
\let\>=\rangle

\let\+=\uparrow

\begin{document}

\thispagestyle{empty}
\vspace*{.5cm}
\noindent
\hspace*{\fill}{CERN-TH/2002-201}\\
\vspace*{1.6cm}

\begin{center}
{\Large\bf Higher-Dimensional Origin of\\[.3cm]
Heavy Sneutrino Domination and\\[.5cm]
Low-Scale Leptogenesis}
\\[2.0cm]
{\large A. Hebecker$^a$, J. March-Russell$^a$, and T. Yanagida$^{a,b}$
}\\[.5cm]
{\it $^a$Theory Division, CERN, CH-1211 Geneva 23, Switzerland}
\\[.2cm]
{\it $^b$Department of Physics, University of Tokyo, Tokyo 113, Japan}
\\[.2cm]
(August 27, 2002)
\\[0.8cm]

{\bf Abstract}\end{center}
\noindent
If the expectation value of the right-handed (rhd) sneutrino comes to 
dominate the universe, its decay naturally leads to successful leptogenesis, 
as well as significant dilution of dangerous inflationary relics, such as 
the gravitino.   The resulting baryon asymmetry is independent of other 
cosmological initial conditions.  This attractive variant of
leptogenesis requires at least one of the rhd neutrinos
to have small Yukawa coupling and to have mass $\sim 10^{6}$ GeV,
much smaller than the grand unified (GUT) scale.
We show that these features naturally arise in the context of independently
motivated and successful 5d orbifold GUTs with inverse-GUT-scale-sized
extra dimensions.  Rhd neutrinos are realized as bulk fields $N_i$ with
5d bulk masses, while Yukawa couplings and lepton-number-violating masses
for the $N_i$ are localized at the SM boundary.  The exponential 
suppression of the would-be $N_i$ zero-modes leads to the desired small
4d Yukawa couplings and small masses for the rhd neutrino states. The see-saw 
prediction for the lhd neutrino mass scale is automatically maintained. 
We show that this realization of rhd neutrinos is nicely accommodated within
an attractive orbifold-GUT flavour model, where all flavour hierarchies
have a geometrical origin.

\newpage

\setcounter{page}{1}

\section{Basics of the scenario}\label{ba}

With growing experimental evidence for neutrino masses in a range that
is consistent with a GUT-scale-based see-saw mechanism~\cite{ss}, 
leptogenesis has become the standard scenario for the generation of the 
baryon asymmetry of the universe. In the original proposal~\cite{fy},
the heavy rhd neutrinos decay in an out-of-equilibrium fashion
once the universe has cooled to a temperature below their mass scale. The 
resulting lepton number is then converted to baryon number by standard 
model (SM) sphaleron processes (for a recent review see, e.g.,~\cite{bp}). 
Alternatively, in a supersymmetric theory, lepton number can be
generated by the decay of a condensate of the scalar component $\tilde{N}$ 
of the rhd neutrino superfield $N$~\cite{my}.  In particular, a recent 
detailed analysis~\cite{hmy} of this scenario has shown that, if $\tilde{N}$
comes to dominate the universe, its decay can naturally produce the required 
lepton asymmetry independently of other cosmological initial conditions.  
At the same time, the number density of dangerous inflationary relics, such 
as the gravitino, is significantly diluted. 

This cosmologically attractive variant of leptogenesis requires at least 
one of the rhd neutrino masses to be very small compared with the GUT scale, 
$\sim 10^{6}\gev$, and the corresponding Yukawa coupling to be suppressed.  
In the present paper, we show that such a situation arises 
naturally in the context of independently motivated higher-dimensional 
GUTs (with inverse-GUT-scale-sized extra dimensions). Before embarking, in 
Sects.~\ref{fm} and~\ref{cos}, upon a detailed discussion of the 
orbifold-GUT model and of the cosmology, we now explain the fundamentals of 
our scenario.

To be specific, we will formulate our ideas in the framework of
supersymmetric SU(5) orbifold GUTs~\cite{kaw,af,hn,hmr} in 5 dimensions.
These theories are attractive because they incorporate the success of MSSM 
gauge-coupling unification~\cite{hn,hmr}, while providing natural 
doublet-triplet splitting as well as suppressed proton decay~\cite{kaw,af,
hn,hmr}.  Moreover, a three generation model with a geometrical origin 
of hierarchical Yukawa couplings and see-saw neutrinos can easily be 
realized~\cite{hmr1}. 

The starting point for these models is a 5d super Yang-Mills theory 
compactified on an interval with coordinate $y\in[0,l]$. At the $y=0$ 
boundary (the `SU(5) brane') the 5d gauge symmetry is unbroken by boundary 
conditions, while at $y=l$ (the `SM brane'), the boundary conditions on the 
gauge fields explicitly break the 5d SU(5) down to the SM gauge group.  
Below the compactification scale $M_c\equiv 1/l$, one has a 4d effective
field theory with SM gauge group (and N=1 supersymmetry).

A basic property of such models is that the `bulk', $y\in(0,l)$, is 
moderately large compared to the fundamental 5d Planck (or UV cutoff)
length. Arguments pointing to this conclusion include the weakness 
of the effective 4d unified coupling, the `observed' smallness of GUT-scale 
threshold corrections, and the flavour hierarchies among the generations 
(see e.g.~\cite{hn,hmr,hmros}). As we discuss in Sect.~\ref{fm}, 
an alternative way of quantifying the size of the bulk follows from the 
requirement that gauge coupling unification (including the KK-mode corrected
logarithmic running above $M_c$) occurs at the 5d Planck scale. Concretely 
this argument favours an orbifold GUT setup with $M l\simeq 300$ and 
$M\simeq 1.4\times 10^{17}\gev$, where $M$ is the reduced 5d Planck mass.
This is in accord with the size of the bulk deduced from other arguments. 

If a bulk mass $m$, odd under 5d parity, is introduced for a bulk matter 
field, its zero mode develops an exponential profile $\sim\exp[-ym 
]$~\cite{bm} (see also~\cite{bf,hmros} and~\cite{exp}).  Thus, depending on 
the sign of $m$, zero modes can be strongly peaked at either brane.  If one 
of the three SM-singlet rhd neutrino fields $N_i$ is exponentially peaked at 
the SU(5) brane, while lepton number violating mass terms $\sim N_i^2$ are 
allowed only at the SM brane, an exponential suppression of both the 4d rhd 
neutrino mass and Yukawa coupling naturally arises.  The crucial 
observation is that the light neutrino masses are not affected even if one 
or more of the $N_i$ are such bulk fields with arbitrary bulk profile.  This 
is clear because the $N_i$ zero modes receive only their kinetic 
term from the bulk, while their effective 4d mass and Yukawa coupling come 
from the brane.  When the $N_i$ are integrated out, their kinetic term plays 
no role and thus it is irrelevant whether they are brane or bulk fields. 
Therefore the traditional see-saw prediction for the lhd neutrino mass scale 
is maintained. However, the rhd sneutrino mass scale is exponentially 
suppressed, as are its Yukawa couplings, and thus decay width.  These are
the new features that allow us to realise the attractive scenario of 
sneutrino ($\tilde{N}$) dominated cosmology and leptogenesis.  

In Sect.~\ref{fm} we present a more detailed motivation and quantitative 
analysis of the basic orbifold picture of neutrino masses and interactions, 
in particular a demonstration that it can be successfully embedded
in a full flavour model.  Specifically, both Higgs doublets and the 
three ${\bf\bar{5}}$'s of SU(5) (denoted by $\bar{F}_i$) are localized at 
the SM brane, while the three ${\bf 10}$'s (denoted by $T_i$) are bulk 
fields. The quark and lepton mass hierarchies are generated by the bulk 
profiles of the $T_i$\,s. 

However, we emphasise that our neutrino mass construction is quite generic 
and does not rely on the details of the specific SU(5) model worked out in 
the rest of this paper. The crucial ingredient is a 5d, or higher-dimensional 
theory compactified on an interval with Yukawa couplings and 
lepton-number-violating neutrino masses localized at one of the boundaries.  
The exponential suppression of zero-mode wave functions at that boundary 
generates both the fermion mass hierarchy and the desired light rhd 
neutrinos. 

The cosmology of the above model of neutrinos has many attractive aspects.  
In particular, over a wide parameter region it leads to the $\tilde{N}$ 
dominated early universe of~\cite{hmy}, as $\tilde{N}$ has an exponentially 
enhanced life-time. In more detail, if the initial value of $|\tilde{N}|$ is 
of the order of $M$, a natural circumstance, then $\tilde{N}$ will 
come to dominate the universe for inflationary reheating temperatures
$T_R\gsim 10^9\gev$.  Moreover, if $T_R$ is varied between $\sim 10^9\gev$
and $\sim 10^{12}\gev$, the gravitino number density in the late universe
remains fixed at the level corresponding to $T_R\sim 10^9\gev$.  This
attractive feature of $\tilde{N}$ dominated cosmology is due to the entropy 
produced by $\tilde{N}$ decay.  Finally, the decay of $\tilde{N}$ produces 
the lepton-number asymmetry.

This cosmology is a fascinating possibility since most of the important 
physical parameters in the present universe, such as baryon-number 
asymmetry, entropy (and, as we later discuss, even spectrum of density 
fluctuations), are determined by the nature of the scalar partner of 
the lightest rhd neutrino. In Sect.~\ref{cos}, we provide a more 
detailed discussion of this cosmology, while some further possibilities, 
together with our conclusions, are contained in Sect.~\ref{con}.

\section{The flavour model}\label{fm}

Consistency of the orbifold GUT framework requires $M_c=1/l$ to
be significantly smaller than the UV scale $M$ of the 5d gauge
theory. To be more specific, the (reduced) Planck masses in 4d ($\bar{M}_P$)
and in 5d ($M$), are related by
\beq
\bar{M}_P^{\,2}= M^3\,l\,, \qquad\qquad \bar{M}_P=M_P/\sqrt{8\pi}\simeq
2.4\times 10^{18}\,\mbox{GeV}\,,\label{mp}
\eeq
and we demand gauge coupling unification at the fundamental scale $M$. In 
spite of the UV sensitivity of the non-renormalizable 5d theory, the 
differences of inverse SM gauge couplings $\alpha_{ij}=\alpha_i^{-1}-
\alpha_j^{-1}$ $(i=1,2,3)$ continue to run logarithmically above 
$M_c$~\cite{hn,hmr} because these differences are only sensitive to the 
SU(5)-breaking SM brane. In the context of the minimal model of~\cite{hmr}, 
where the Higgs-doublets are localized at the SM brane, this `differential
running'~\cite{nsw} comes entirely from the gauge sector. With the effective 
SUSY breaking scale set to $m_Z$, we have
\beq
\alpha_{ij}(m_Z)=\alpha_{ij}(M)+\frac{1}{2\pi}\left\{a_{ij}\ln\frac{M}{m_Z}+
\frac{1}{2}b_{ij}\ln\frac{M}{M_c}\right\}\,,\label{run}
\eeq
where $a_{ij}=a_i-a_j$ and $b_{ij}=b_i-b_j$ (with $a_i=(33/5,1,-3)$ and 
$b_i=(-10,-6,-4)$) characterise the familiar MSSM running and the KK 
mode contributions respectively.  If we define the conventional 4d
unification scale by the meeting of the U(1) and SU(2) couplings
$\alpha_1$ and $\alpha_2$, then the low-energy data $\alpha_i^{-1}(m_Z) 
=(59.0,29.6,8.4)$ imply $M_{\rm GUT}\simeq 1.9\times 10^{16}\gev$. 
By contrast, combining Eqs.~(\ref{mp}) and (\ref{run}) and assuming
$\alpha_{12}^{-1}(M)=0$, one derives the 5d unification scale $M=1.4\times 
10^{17}\gev$ and $Ml=2.8\times 10^2$. Of course, these numbers represent 
only rough estimates since the $\alpha_{ij}(M)$ have, in general, non-zero 
${\cal O}(1)$ values, which perturb the calculation of $Ml$ and $M$.
(In the slightly different approach of~\cite{hn1}, the model is fixed by 
requiring the precision of simultaneous 1-2 and 2-3 unification to be better 
than with conventional MSSM running.) 

The above discussion provides us with a motivation for an orbifold GUT 
setup with $M\simeq 1.4\times 10^{17}\gev$ and with the small parameter 
$\epsilon^2\equiv 1/(Ml)\simeq 1/300$. Flavour is described by introducing 
the three $\bar{F}_i$ fields and the two Higgs doublets $H_u$ and 
$H_d$ on the SM brane (recall that, because of the reduced symmetry
of the SM brane, there is no need for Higgs triplets), while allowing
the $T_i$ to propagate in the bulk.\footnote{
To 
be more precise, one introduces hypermultiplets $T_i$ and $T_i'$ and 
assigns boundary conditions ensuring that the zero modes correspond to
the field content of three ${\bf 10}$'s of SU(5)~\cite{hn,hmr}. Moreover, 
the boundary conditions break N=2 to N=1 SUSY, leaving us with conventional 
chiral multiplets at low energy $E<M_c$.
}
This large disparity between $T$'s and $\bar{F}$'s is the geometric origin 
of small quark and large lepton mixing (cf.~\cite{wy} and~\cite{hmr1}). 
If a bulk mass $m$, odd under 5d parity, is introduced for a 5d 
hypermultiplet, its zero mode develops an exponential profile 
$\sim\exp[-ym]$. Thus, depending on the sign of $m$, zero modes can be 
strongly peaked at either brane. In particular, this allows for a
dynamical realization of SM-brane fields with quantum numbers
appropriate for an SU(5) representation (e.g., the Higgs doublets
and the $\bar{F}_i$ above). We will use this additional tool to realize the 
fermion mass hierarchy by appropriately localizing the $T_i$. 

An understanding of the observed hierarchies in the fermion masses and 
mixings emerges naturally if the bulk mass of $T_3$ is sufficiently large 
and negative, $m<0$ (so that, for all practical purposes, $T_3$ is a SM brane 
field), while $T_2$ has vanishing bulk mass (flat zero-mode) and $T_1$ has 
a finite bulk mass $m>0$ (its zero-mode therefore being suppressed at the 
SM brane). 

Concretely, the primordial and unstructured ${\cal O}(1)$ Yukawa couplings 
$\lambda$ at the SM brane are rescaled as
\beq
\lambda\to\frac{\lambda}{\sqrt{Ml}}\,\sqrt{\frac{2ml}{e^{2ml}-1}}
\eeq
for each participating bulk field with bulk mass $m$. This rescaling 
follows from the 4d canonical normalization of the 5d kinetic term and the 
exponentially suppressed zero-mode field value at the SM brane. Applying 
this to $T_2$, one finds that this field enters Yukawa interactions with
a suppression factor $\epsilon$. The analogous suppression factor for
$T_1$ depends on $m$ and becomes $\sim\epsilon^2$ for the choice $ml\simeq
3.9$. This leads to the following realistic Yukawa matrix structure for the
two effective 4d interactions $H_uT^T\lambda_{TT}T$ and $H_dT^T\lambda_{TF} 
\bar{F}$:
\beq
\lambda_{TT}\sim\left(\begin{array}{ccc}
\epsilon^4 & \epsilon^3 & \epsilon^2\\
\epsilon^3 & \epsilon^2 & \epsilon\\
\epsilon^2 & \epsilon & 1 
\end{array}\right)\,\,,\qquad\lambda_{TF}\sim\left(\begin{array}{ccc}
\epsilon^2 & \epsilon^2 & \epsilon^2\\
\epsilon & \epsilon & \epsilon\\
1 & 1 & 1 
\end{array}\right)\,\,,
\eeq
with unknown ${\cal O}(1)$ factors multiplying each entry.\footnote{
A 
slight modification, leading to a welcome further suppression of electron 
and down-quark mass, is obtained by placing one of the $\bar{F}$'s in
the bulk. In fact, such a construction can be motivated by its particularly
high symmetry: One set of fields $(T,\bar{F})$ are on the SM brane, one set
are massless bulk fields, and the third set are massive bulk fields with the 
sign of the mass flipped between $T$ and $\bar{F}$ (making $\bar{F}$
effectively a SM brane field).
} 
It is known that this Yukawa coupling hierarchy also gives rise to an 
approximately correct CKM structure. The top Yukawa coupling is naturally 
${\cal O}(1)$. The required relative suppression of down-type Yukawa 
couplings can be realized either by going to large tan$\beta$ or by slightly 
decreasing the strength with which $H_d$ is peaked at the SM brane. 

The construction presented so far can be summarized as follows. By 
identifying the 5d Planck mass with the unification scale, we have argued 
for a relative bulk size characterized by $\epsilon\sim 1/\sqrt{Ml}\sim 1/ 
\sqrt{300}$. If all fields except $T_1$ and $T_2$ are localizing at the SM 
brane, this bulk suppression factor beautifully explains the mass hierarchy 
between the two heavier generations~\cite{hmros}. To explain the extreme 
lightness of the first generation, we had to give $T_1$ a bulk profile 
exponentially suppressed at the SM brane using the additional tool of bulk 
masses. With this tool in hand, rhd neutrino singlets can easily acquire the 
exponentially suppressed 4d masses and couplings required for the 
$\tilde{N}$ dominated universe. 

Now we discuss the rhd neutrinos in more detail. Consider introducing 
three neutrino fields $N_i$ at the SM brane. Given a Majorana 
mass matrix $M_{N,ij}$ with ${\cal O}(M)$ entries and ${\cal O}(1)$ Yukawa 
couplings between $N_i$, $\bar{F}_i$ and $H_u$, the conventional see-saw 
mechanism leads to a light neutrino mass scale $|H_u|^2/M\simeq 2\times 
10^{-4}$ eV. In the present scenario, such a small mass scale is welcome 
since it ensures the out-of-equilibrium decay of $\tilde{N}$ (see 
Sect.~\ref{cos}). The observed neutrino oscillations, which require a 
somewhat larger light neutrino mass scale, can be accommodated by
assuming that $M_N$ has two slightly suppressed eigenvalues. (A concrete
example of such a suppression mechanism will be provided shortly.)

As discussed in Sect.~\ref{ba}, light neutrino masses are not affected if 
one or more of the $N_i$ are promoted to bulk fields. To be specific, let us 
declare $N_1$ to be a bulk field with bulk mass $m_1$ and effective 4d mass 
(cf.~Sect.~5 of~\cite{hmr1})
\beq
M_1\simeq 2m_1 e^{-2m_1l}\,.\label{ems}
\eeq
Due to the exponential suppression factor, the desired small value of $M_1$ 
is easily realized: for example, $M_1\simeq 3\times 10^{6}\gev$ for $m_1l 
\simeq 11$. While this concludes the description of our basic flavour model 
with a naturally light rhd neutrino, several open issues deserve further
discussion.

Firstly, we need to enhance two of the light neutrino masses. For example, 
one could introduce a Froggatt-Nielsen U(1), broken by two charge-($\pm$1) 
fields with vacuum expectation values $|\Phi_\pm|$ where $|\Phi_\pm|/M 
\simeq\eta\ll 1$~\cite{fn}. With charge assignments (0,$-$1,$-$1) and 
(1,1,1) for the $N_i$ and $\bar{F}_i$ respectively, one obtains the 
following structures for the Yukawa matrix $\lambda_N$ in $H_u\bar{F}_i^T 
\lambda_{N,ij}N_j$ and the mass matrix:
\beq
\lambda_N\sim\left(\begin{array}{ccc}
\eta & 1 & 1\\
\eta & 1 & 1\\
\eta & 1 & 1
\end{array}\right)\,\,,\qquad M_N\sim\left(\begin{array}{ccc}
1 & \eta & \eta\\
\eta & \eta^2 & \eta^2\\
\eta & \eta^2 & \eta^2 
\end{array}\right)\,\,.
\eeq
It is easy to convince oneself that all entries of the resulting light 
neutrino mass matrix $m_\nu\simeq \lambda_NM_N^{-1}\lambda_N^T\times |H_u|^2$ 
are of the order $\eta^{-2}|H_u|^2/M$. This also sets the scale for two
of the eigenvalues. Although the remaining eigenvalue is suppressed to
$\eta^2 |H_u|^2/M$, all three mixing angles are generically large.\footnote{
The 
smallness of 1-3 mixing may be accidental (cf.~\cite{anarchy}). 
Alternatively, it could be explained in a modified model where one of
the $\bar{F}$'s is a bulk field.
}
In our setup, realistic neutrino phenomenology requires $\eta\sim 10^{-1}$. 
Furthermore, assigning a suitable U(1) charge to $H_d$ provides an 
alternative way to realize suppressed down-type masses. Let us finally argue 
why the family-symmetry should be broken in the U(1) charge assignment of the 
$N_i$. One possibility is to demand vanishing U(1) charges for all bulk 
fields. Alternatively, one could replace the U(1) with a $Z_3$ and then note 
that, while the cancellation of the mixed $Z_3$-SU(5) anomaly forces all three
$\bar{F_i}$ to have the same charge, the $N_i$ charges remain unrestricted. 

Secondly, it is necessary to forbid both parity-even bulk masses as well 
as SU(5)-brane-localized mass terms for the rhd neutrinos. 
Following~\cite{hmr1}, this can be done by gauging U(1)$_\chi$ (named as 
in~\cite{pdg}), defined by SU(5)$\times$U(1)$_\chi\subset\,$SO(10). Since 
$\tilde{N}$ domination requires a large initial value of $\tilde{N}_1$, the 
$D$-term potential for $\tilde{N}_1$ has to be suppressed. This can be 
achieved by dynamically breaking U(1)$_\chi$ at the high scale $M$ in
5d. A surviving discrete subgroup will be sufficient to forbid the dangerous
mass operators. In addition, it is natural that U(1)$_\chi$ is broken by
orbifolding at the SM brane~\cite{hmr1}, making it the only possible 
location for the required lepton-number violating mass term.

Furthermore, we would like to comment on the relation of our method of 
generating the fermion mass hierarchy and the light $N_1$ field to the 
familiar Froggatt-Nielsen approach. Certainly the assignment of bulk masses 
to different sets of fields resembles the assignment of U(1) charges. This 
similarity becomes even more pronounced if the bulk masses are dynamically 
realized by expectation values of U(1) fields with Fayet-Iliopoulos terms 
at the boundary (see, e.g.,~\cite{fi}). However, especially in the case of 
the large suppression factor needed for $N_1$, it is a significant advantage 
that the bulk mass effect is exponential rather than power-like. 
Furthermore, there are crucial qualitative differences in the resulting 
phenomenology. For example, higher-order K\"ahler-terms involving 
$T^\dagger_iT_i$ together with the SUSY-breaking spurion, which can lead to 
dangerous flavour violation, are unrestricted by U(1) symmetries. In our 
case, if SUSY breaking is localized at the SM brane, such terms will be 
geometrically suppressed for the first two generations. The argument 
extends to the $\bar{F}_i$ if some of them are promoted to bulk fields. 

Finally, note that the bulk masses used in the above construction are 
significantly smaller than the fundamental 5d scale $M$. This may follow
naturally if bulk masses come from expectation values of weakly coupled 
U(1) fields. Alternatively, one may imagine the 5d theory to descend
from a 6d theory, where bulk masses are forbidden, so that 5d bulk masses
are due to small, non-perturbative effects arising in the 6d to 5d 
compactification process.

\section{Cosmology}\label{cos}

Let us turn to the discussion of cosmology. 
It is a reasonable assumption that the scalar partner of at least
one of the right-handed neutrinos has, during inflation, an expectation
value of the order of the cutoff scale $M$.  The reason for this
is that higher-dimension operators in the K\"ahler potential link the
inflationary sector, in particular the superfield whose $F$-term or $D$-term
gives rise to the non-zero vacuum energy, and the rhd neutrino superfields.
(Higher order superpotential terms can be forbidden by a continuous or 
discrete symmetry acting on the superfield $N_1$.) This leads to a 
contribution to the $({\rm mass})^2$ of the sneutrino of order $\meff^2 
\sim(H_{\rm inf})^2$, where $H_{\rm inf}$ is the inflationary expansion 
rate. The sign depends upon the unknown K\"ahler operator coefficient. If 
$\meff^2<0$, then 
$\tilde{N}_1$ gains an expectation value only limited by yet higher-order 
terms in the K\"ahler potential, suppressed by powers of $M$. 
(Note that the parametrically small inflationary $F$- or $D$-term 
expectation value multiplies the entire set of higher dimension operators 
which lead to a potential for $\tilde{N}_1$.) Here we have assumed that 
$H_{\rm inf}$ is larger than the mass of the right-handed neutrino.

As described in the previous sections, for the lightest rhd sneutrino field 
$\tilde{N}_1$ both its mass $M_1$ (cf.~Eq.~\ref{ems}) and its effective 4d
Yukawa coupling, 
\beq
\lambda_{N,i1}\simeq \eta \sqrt{2m_1/M}\,e^{-m_1l}\,, \label{yc1}
\eeq
are exponentially suppressed. When the expansion rate $H$ after the end of 
inflation decreases below $M_1$, $\tilde{N}_1$ starts coherently oscillating. 
Given a condition on the post-inflationary reheating temperature $T_R$ (to be 
discussed below), the oscillation energy dominates the energy density of the
early universe, and its decay produces the baryon asymmetry observed today 
without any cosmological difficulty.
  
If the Yukawa 
couplings $\lambda_{N,ij}$ of the $N_j$ have $CP$ violating phases, the 
decay of $\tilde{N}_1$ produces a lepton-number asymmetry $\epsilon_1$ 
given by~\cite{by,hmy}
\beq
\epsilon_1 \simeq 1\times
10^{-10}(M_1/10^6\gev)(m_{\nu_3}/0.05\ev)\delta_{eff}. \label{lepton}
\eeq
Here, $\delta_{eff}$ is an effective $CP$ violating phase. This lepton
asymmetry is converted into a combined baryon and lepton asymmetry through
non-perturbative electroweak sphaleron effects.   
A crucial observation of ref.~\cite{hmy} is that the final baryon asymmetry 
is determined by the reheating temperature, $T_{N_1}$, of the $\tilde{N}_1$ 
decay once it dominates the energy density of the early universe.  The 
net baryon to entropy ratio is given by~\cite{hmy}
\beq
n_B/s \simeq (8/23)(3/4)\epsilon_1(T_{N_1}/M_1)\simeq 0.3\times 10^{-10} 
(T_{N_1}/10^6\gev)(m_{\nu_3}/0.05\ev)\delta_{eff} \label{baryon},
\eeq
where $n_B$ and $s$ are baryon-number and entropy densities, respectively. 
The observed baryon asymmetry $n_B/s\simeq (0.4-1)\times 10^{-10}$ is 
obtained by taking $T_{N_1}\simeq 2\times 10^6\gev$ for $\delta_{eff}\simeq 
1$. On the other hand, the reheating temperature $T_{N_1}$ due to
$\tilde{N}_1$ decay is given by
$T_{N_1}^2\simeq \Gamma_{N_1}\,\bar{M}_P$, where the decay rate
$\Gamma_{N_1}$ is
\beq
\Gamma_{N_1}\simeq (3/4\pi)\lambda_{N,i1}^2 M_1. \label{decay-rate}
\eeq
We see that the desired reheating temperature $T_{N_1}\simeq 2\times 10^6\gev$ 
is obtained for $m_1l=11$, where we have used Eqs.~(\ref{ems}) and 
(\ref{yc1}).  Notice that the rhd neutrino mass $M_1\simeq 3\times 10^6 
\gev > T_{N_1}$, and hence the out-of-equilibrium condition for 
$\tilde{N}_1$ decay is automatically satisfied.

Let us now discuss the condition for the $\tilde{N}_1$ domination in the early
universe.  Since the initial value of $|\tilde{N}_1| \simeq 10^{17}\gev$, the
energy density of the coherent $\tilde{N}_1$ oscillation, at the start of
this oscillation, is a minor component of the total density.  However, if the
$\tilde{N}_1$ lifetime is sufficiently longer than that of the 
inflaton, it can dominate
the early universe since the energy density of the radiation resulting
from the inflaton decay dilutes faster than the energy density of the 
coherent oscillation.  Thus, the condition for the $\tilde{N}_1$ domination 
is translated to an upper limit on the inflaton lifetime for a given 
$\tilde{N}_1$ lifetime.  Written in terms of the post-inflationary
reheating temperature, $T_R$, this condition is 
\beq
T_R > 3T_{N_1}(\bar{M}_P/|\tilde{N}_1|)^2
\simeq 2\times 10^9 \gev\,, \label{domination}
\eeq
which is easily satisfied in a variety of inflationary models.

The post-inflationary reheating temperature must also satisfy 
an upper bound so as to avoid the 
over-production of gravitinos.  In the standard cosmology (without 
$\tilde{N}_1$ domination) the upper bound is determined to be
$T_R<10^{10}\gev$ for a gravitino mass $\sim 1\tev$~\cite{grav}
(for a recent analysis see also~\cite{kkm}), giving a stringent
restriction on inflationary models. 

However, with $\tilde{N}$-dominated cosmology, only a weaker constraint
applies.  The reason for this is that
$\tilde{N}_1$ decay reheats the universe once more, and the associated
entropy production dilutes substantially the density of earlier-produced
gravitinos.  In detail, start from a situation where $T_R$ is at its lower 
bound given by Eq.~(\ref{domination}) and raise the reheating temperature 
gradually. While the post-inflationary gravitino production increases 
proportionally to $T_R$, the subsequent $\tilde{N}_1$ decay introduces the 
dilution factor
\beq
1/\Delta \simeq 3T_{N_1}/T_R(\bar{M}_P/|\tilde{N}_1|)^2\,, \label{dilution}
\eeq
which precisely compensates the previous effect. Thus, the number density 
of gravitinos is determined by an effective reheating temperature 
$T_{R,eff}$ instead of the original $T_R$. This effective temperature is 
given by
\beq
T_{R,eff} = 1/\Delta\times T_R = 3T_{N_1}(\bar{M}_P/|\tilde{N}_1|)^2 
\simeq 2\times 10^9 \gev.\label{eff-T}
\eeq
However, if the reheating temperature rises above $T_R\simeq 10^{12}$ GeV, 
the situation changes because now reheating takes place before $\tilde{N_1}$ 
oscillations start. While the initial gravitino production continues to grow 
with $T_R$, the dilution factor remains constant, giving
\beq
T_{R,eff} = 1/\Delta\times T_R = 3T_{N_1}(\bar{M}_P/|\tilde{N}_1|)^2
(T_R/10^{12}\gev) 
\simeq 2\times 10^9 \gev (T_R/10^{12}\gev) .\label{eff-T2}
\eeq

Applying the analysis of Ref.\cite{grav}, we see that a gravitino
of mass of order $1\tev$ is consistent with a significantly extended
range of post-inflationary reheating temperatures, $T_R<10^{13} \gev$.
Putting this upper bound together with our earlier lower bound from
$\tilde{N}$ domination leads to an allowed range, $2\times 10^9\gev<T_R< 
10^{13} \gev$, for successful sneutrino-dominated leptogenesis.

\section{Conclusions}\label{con}
In summary, we have presented a higher-dimensional scenario, well motivated
from a particle-physics perspective, in which cosmological heavy sneutrino
domination occurs naturally and low-scale leptogenesis is responsible for the 
observed baryon asymmetry of the universe. The entropy produced in the decay 
of the $\tilde{N}$ condensate dilutes unwanted relics from the period of 
reheating, alleviating in particular the danger of gravitino over-production. 

We briefly comment on density perturbations in this scenario. The dominant 
density perturbation can originate from the fluctuations of $\tilde{N}$ 
during inflation~\cite{ly}. The deviation of the spectrum from scale 
invariance depends upon $\meff^2/H_{\rm inf}^2$, where $\meff^2$ is the 
effective (mass)$^2$ discussed at the beginning of Sect.~\ref{cos}.
Scale invariance thus requires the dimensionless K\"ahler potential couplings 
between $N$ and the inflaton to be slightly suppressed. Moreover, as in 
our scenario $\tilde{N}$ decay is the origin of baryon number asymmetry 
and dark matter, we necessarily have adiabatic perturbation 
dominance~\cite{mo}. 

Thus, the role of the inflaton is reduced to providing a period of 
exponential expansion while its two main dynamical effects, the production 
of density perturbations and the reheating of the universe, are taken over 
by the heavy sneutrino. In fact, the presence of the inflaton mainly has a 
constraining effect -- it has to decay sufficiently early to allow for heavy 
sneutrino domination and sufficiently late not to produce an excess of
gravitinos. Thus, one might also wonder whether it is possible to get rid 
of the inflaton altogether. One obvious possibility would be to assume a 
sneutrino potential with a flat region away from the origin, so that a 
sufficiently long inflationary period driven by the sneutrino condensate is 
realized. It would be a very interesting and challenging task to understand 
the origin of such an unusual sneutrino potential. 

Finally, we discuss a less radical way of avoiding the constraints 
associated with the decay of the inflaton entirely. If the inflaton 
potential is such that, in the true vacuum, the inflaton is massless,
its energy density during the oscillation period decays faster\footnote{
We 
are grateful to J. Garcia-Bellido for helpful discussions on this point.
}
than that of $\tilde{N}$, which varies with the scale factor $R$ as 
$R^{-3}$. Thus, the desired $\tilde{N}$ domination is always obtained. The 
masslessness of the inflaton is technically natural since we do not 
require any non-gravitational coupling of the inflaton to the matter sector. 
It would be interesting to write down and analyse a well-motivated and 
complete inflation model with a potential that leads to such a `harmless' 
late time behaviour of the inflaton. 

\noindent
{\bf Acknowledgements}: We thank David Lyth for
helpful discussions. The work by T.Y. is supported in part by 
Grant-in-Aid for Scientific Research (S)14102004.


\begin{thebibliography}{99}

\bibitem{ss} 
T. Yanagida, in Proceedings of the {\it Workshop on Unified Theories and
Baryon Number of the Universe}, editors; O.Sawada and A.Sugamoto,
(Tsukuba, Japan, 1979, KEK Report KEK-79-18) p. 95;\\
M. Gell-Mann, P. Ramond, and R. Slansky, in {\it Supergravity} (North 
Holland, Amsterdam, 1979) eds. P. van Nieuwenhuizen and D. Freedman, p. 315.

\bibitem{fy}
M.~Fukugita and T.~Yanagida, Phys.\ Lett.\ B {\bf 174} (1986) 45.

\bibitem{bp}
W.~Buchm\"uller and M.~Pl\"umacher, Int.\ J.\ Mod.\ Phys.\ A {\bf 15}
(2000) 
5047 \\{} [arXiv:hep-ph/0007176].

\bibitem{my}
H.~Murayama and T.~Yanagida, Phys.\ Lett.\ B {\bf 322} (1994) 349\\{}
[arXiv:hep-ph/9310297].

\bibitem{hmy}
K.~Hamaguchi, H.~Murayama and T.~Yanagida, Phys.\ Rev.\ D {\bf 65}
(2002) 
043512 [arXiv:hep-ph/0109030].

\bibitem{kaw}
Y.~Kawamura,
Prog.\ Theor.\ Phys.\  {\bf 105} (2001) 999
[arXiv:hep-ph/0012125].

\bibitem{af}
G.~Altarelli and F.~Feruglio, Phys.\ Lett.\ B {\bf 511} (2001) 257
[arXiv:hep-ph/0102301].

\bibitem{hn}
L.~J.~Hall and Y.~Nomura, Phys.\ Rev.\ D {\bf 64} (2001) 055003
[arXiv:hep-ph/0103125].

\bibitem{hmr}  
A.~Hebecker and J.~March-Russell, Nucl.\ Phys.\ B {\bf 613} (2001) 3\\{}
[arXiv:hep-ph/0106166];
Nucl.\ Phys.\ B {\bf 625} (2002) 128
[arXiv:hep-ph/0107039].

\bibitem{hmr1}
A.~Hebecker and J.~March-Russell, Phys.\ Lett.\ B {\bf 541} (2002) 338\\{}
[arXiv:hep-ph/0205143].

\bibitem{hmros}
L.~Hall, J.~March-Russell, T.~Okui and D.~R.~Smith,
arXiv:hep-ph/0108161;\\
L.~J.~Hall, Y.~Nomura and D.~R.~Smith, arXiv:hep-ph/0107331.

\bibitem{bm}   
R.~Jackiw and C.~Rebbi, Phys.\ Rev.\ D {\bf 13} (1976) 3398;\\
D.~B.~Kaplan, Phys.\ Lett.\ B {\bf 288} (1992) 342
[arXiv:hep-lat/9206013].

\bibitem{bf}
N.~Arkani-Hamed \etal, Phys.\ Rev.\ D {\bf 65} (2002) 024032
[arXiv:hep-ph/9811448];
N.~Arkani-Hamed and S.~Dimopoulos, Phys.\ Rev.\ D {\bf 65} (2002)
052003\\{}
[arXiv:hep-ph/9811353].

\bibitem{exp}   
H.~Georgi, A.~K.~Grant and G.~Hailu, Phys.\ Rev.\ D {\bf 63} (2001)
064027
\\{} [arXiv:hep-ph/0007350];\\
D.~E.~Kaplan and T.~M.~Tait, JHEP {\bf 0111} (2001) 051 
[arXiv:hep-ph/0110126].

\bibitem{nsw}  
Y.~Nomura, D.~R.~Smith and N.~Weiner, Nucl.\ Phys.\ B {\bf 613} (2001) 
147\\{} [arXiv:hep-ph/0104041].

\bibitem{hn1}
L.~J.~Hall and Y.~Nomura,
Phys.\ Rev.\ D {\bf 65} (2002) 125012 [arXiv:hep-ph/0111068].

\bibitem{wy} T.~Watari and T.~Yanagida, arXiv:hep-ph/0205090.

\bibitem{fn}
C.~D.~Froggatt and H.~B.~Nielsen, Nucl.\ Phys.\ B {\bf 147} (1979) 277.

\bibitem{anarchy}
L.~J.~Hall, H.~Murayama and N.~Weiner, Phys.\ Rev.\ Lett.\  {\bf 84}
(2000) 
2572\\{} [arXiv:hep-ph/9911341].

\bibitem{pdg}
D.~E.~Groom {\it et al.}  [Particle Data Group],
Eur.\ Phys.\ J.\ C {\bf 15} (2000)~1 (see p. 290).

\bibitem{fi}
R.~Barbieri \etal, arXiv:hep-th/0203039;\\
S.~Groot Nibbelink, H.~P.~Nilles and M.~Olechowski,
arXiv:hep-th/0205012;\\
D.~Marti and A.~Pomarol, arXiv:hep-ph/0205034.

\bibitem{by}
W.~Buchm\"uller and T.~Yanagida, Phys.\ Lett.\ B {\bf 445} (1999) 399\\{}
[arXiv:hep-ph/9810308].

\bibitem{grav}
M. Y. Khlopov and A. D. Linde, Phys. Lett. B138, 265 (1984);\\
J. Ellis, J. E. Kim and D. V. Nanopoulos, Phys. Lett. B145, 181 (1984);\\
M. Kawasaki and T. Moroi, Prog. Theor. Phys. 93, 879 (1995)\\{}
[arXiv:hep-ph/9403364].

\bibitem{kkm}
M. Kawasaki, K. Kohri and T. Moroi, Phys. Rev. D63, 103502 (2001)\\{}
[arXiv:hep-ph/0012279].

\bibitem{ly}
D.~H.~Lyth and D.~Wands, Phys.\ Lett.\ B {\bf 524} (2002) 5
[arXiv:hep-ph/0110002];\\
T.~Moroi and T.~Takahashi, Phys.\ Lett.\ B {\bf 522} (2001) 215
[Erratum-ibid.\ B {\bf 539} (2002) 303] [arXiv:hep-ph/0110096].

\bibitem{mo}
T.~Moroi and T.~Takahashi, arXiv:hep-ph/0206026.

\end{thebibliography}
\end{document}